\documentclass[reprint,amsmath,amssymb,aps,floats,superscriptaddress,preprintnumbers]{revtex4-2}

\usepackage{graphicx}
\usepackage{dcolumn}
\usepackage{bm}
\usepackage{hyperref}

\usepackage{color}
\usepackage{cancel}
\usepackage[normalem]{ulem}
\usepackage{AAS_macro}
\usepackage{multirow}
\usepackage{orcidlink}

\usepackage{amsmath}	
\usepackage{amssymb}	
\usepackage{mathtools}

\renewcommand{\arraystretch}{1.4}

\begin{document}

\title{Cosmological constraints from Type I radio-loud quasars}

\author{L. Huang\orcidlink{0000-0003-4545-7066}}
\email{huanglong20122021@163.com}
\affiliation{College of Science, Jiujiang University, Jiujiang 332000, People's Republic of China.}
\affiliation{Key Laboratory of Functional Microscale Materials in Jiangxi Province,\\
 Jiujiang 332000, People's Republic of China.}

\author{ Z. Y. Tu}
\affiliation{College of Science, Jiujiang University, Jiujiang 332000, People's Republic of China.}
\affiliation{Key Laboratory of Functional Microscale Materials in Jiangxi Province,\\
 Jiujiang 332000, People's Republic of China.}

\author{ N. Chang}
\affiliation{Xinjiang Astronomical Observatory, Chinese Academy of Sciences, \\
Urumqi 830011, People's Republic of China.}
\affiliation{Key Laboratory of Radio Astronomy, Chinese Academy of Sciences, \\
Nanjing, 210008, People's Republic of China.}

 \author{ F. F. Song}
\affiliation{College of Science, Jiujiang University, Jiujiang 332000, People's Republic of China.}
\affiliation{Key Laboratory of Functional Microscale Materials in Jiangxi Province,\\
 Jiujiang 332000, People's Republic of China.}

 \author{ F. He}
\affiliation{Institute of Physics, Hunan University of Science and Technology, \\
Xiangtan, Hunan 411201, People's Republic of China.}

 \author{ X. Y. Fu}
\affiliation{Institute of Physics, Hunan University of Science and Technology, \\
Xiangtan, Hunan 411201, People's Republic of China.}

\begin{abstract}

We obtain a new sample of 1192 Type I quasars with the UV-optical, radio and X-ray wavebands coverage by combining \citet{Huang2022} and other matching data of SDSS-DR16 with FIRST, XMM–Newton, and Chandra Source Catalog, and a sample of 407 flat-spectrum radio-loud quasars (FSRLQs) of blazars from the Roma-BZCAT, which can be used to investigate their multi-band luminosity correlations and measure the luminosity distances of these Type I radio-loud quasars (RLQs) samples. We check the correlation between X-ray, UV-optical, and radio luminosity for various groupings of radio-quiet quasars (RQQs) and RLQs by parameterizing X-ray luminosity as a sole function of UV-optical or radio luminosity and as a joint function of UV/optical radio luminosity, which also can be employed to determine these cosmological distances. By Bayesian information criterion (BIC), the data suggest that the X-ray luminosity of RQQs is indirectly correlative with radio luminosity because of the connection between UV-optical and radio luminosity. But for RLQs, the X-Ray luminosity is directly related to radio luminosity, and the correlations between X-ray, optical/UV, and radio luminosity increase with the ratio of monochromatic luminosities logR. Meanwhile, we compare the results from RLQs with different UV-optical power law index ${\Gamma _{UV}}$, the goodness of fit for RLQs with ${\Gamma _{UV}}\le 1.6$ seems to be better. Finally, we apply a combination of Type I RLQs and SN Ia Pantheon to verify the nature of dark energy concerning whether or not its density deviates from the constant, and give the statistical results.

\end{abstract}


\maketitle

\section{Introduction} \label{Sec:1}

A great number of quasars data have been obtained and used to investigate their luminosity correlation. There is a dichotomy in the distribution of the radio luminosity of quasars \citep{Strittmatter1980}, which depends on the ratio of monochromatic luminosities measured at (rest frame) 5 GHz and 2500 {\AA} \citep{Kellermann1989, Stocke1992, Kellermann1994}. RLQs are often defined by $log R > 1$ and RQQs satisfy $log R \le 1$. A large number of data suggest that the X-ray luminosity of RQQs is related to the UV-optical luminosity \citep{Tananbaum1983, Worrall1987, Miller2010, Zhu2020, Browne1987}, which also indicates that X-ray emission is created by Compton upscattering of disk photons occurring in a hot "corona". The X-ray properties of RLQs are different from those of RQQs. The X-ray emission of RLQs are not merely contributed to inverse compton scattering, but also powered directly or indirectly by the radio jet \citep{Evans2006, Hardcastle2009, Miller2010, Huang2022}, which can be verified by parameterization methods.

On the other hand, quasars can also be categorized by whether they have broad emission lines (Type I), only narrow lines (Type II), or no lines except when a variable continuum is in a low phase (Blazars) \citep{Urry1995, Sulentic2000}, and blazars are generally divided into two classes on the basis of their optical spectra. The first class is represented by the flat-spectrum radio-loud quasars (FSRLQs), the second class is the BL Lac objects characterized by featureless spectra with emission/absorption lines of equivalent width lower than 5 {\AA} \citep{DAbrusco2014, DAbrusco2019}.

To investigate the multi-band luminosity correlations of quasars and measure the luminosity distances of these Type I quasars, we construct a large sample of Type I quasars by combining \citet{Huang2022} and other matching data of SDSS-DR16 with FIRST, XMM –Newton, and Chandra Source Catalog, and a sample of 407 FSRLQs of blazars from the Roma-BZCAT. Meanwhile, we compare the X-ray luminosity relation of RLQs with different UV-optical power law index ${\Gamma _{UV}}$ and X-ray photon index ${\Gamma _X}$.

In addition, Worrall et al. have used Type I RLQs to check whether their luminosity correlation depends on redshift \citep{Miller2010}. Hence, we also consider dividing the RLQs sample into different redshift bins, which can be applied for segment fitting and examining whether the X-ray luminosity relation is redshift-dependent.

In Section \ref{Sec:2} of this paper, we introduce the source of data used, including Type I quasars and blazars. In Section \ref{Sec:3}, we adopt three parametric models to analyze the X-ray luminosity correlation of RQQs and RLQs, which include X-ray luminosity as a sole function of UV-optical or radio luminosity and as a joint function of UV-optical and radio luminosity. We compare and analyze the results from three different models by using the Bayesian information criterion (BIC). Furthermore, we subdivide the RLQs sample into various redshift bins, which can be used for testing whether there is a redshift evolution of the X-ray luminosity relation. In Section \ref{Sec:4}, we employ the X-ray luminosity relation of Type I RLQs to measure and obtain their cosmological luminosity distance.  In Section \ref{Sec:5}, we apply a combination of Type I RLQs and SN Ia Pantheon to test the nature of dark energy by reconstructing the dark energy equation of state $w(z)$, which concerns whether or not the density of dark energy evolves with time. In Section \ref{Sec:6}, we summarize this paper.

\section[ data used]{ data used} \label{Sec:2}

Modern optical instruments and surveys (e.g. Sloan Digital Sky Survey; SDSS) \citep{Lyke2020, Ahumada2020, Paris2017, Alam2015, Richards2002}; Radio surveys (e.g. FaintImages of the Radio Sky at Twenty-Centimeters; FIRST) \citep{Helfand2015}, and archival X-ray data from XMM–Newton \citep{Rosen2016, Webb2020}, Chandra \citep{Evans2010}, provide large amounts of quasars data, which can be applied to check the multi-band luminosity correlation for quasars. The 16th data release (DR 16) from the SDSS presented a quasar catalog including the spectra of 750,414 quasars \citep{Lyke2020}, in addition, a catalog containing 946,432 sources observed at a frequency of 1.4GHz were released by FIRST \citep{Helfand2015}.

We first matched the SDSS-DR16 quasar catalogue with the latest FIRST survey data using a $2''$ matching radius, all Type I quasars flagged as broad absorption lines (BALs) are removed, and obtained a matched sample of Type I quasars with the UV-optical and radio wavebands coverage. Next we matched this sample to the latest XMM-Newtom Source Catalog and the Chandra Source Catalog Release 2.0 to obtain their X-ray fluxes (0.2-12 keV for XMM-Newtom and 0.5-7 keV for Chandra) \citep{Rosen2016, Webb2020, Helfand2015}, with a matching radius of $5''$. Finally, we construct a large sample of Type I quasars with multi-wavelength coverage, and some of these objects are from \citet{Huang2022}.

For this new sample, the UV-optical power-law index ${\Gamma _{UV}}$ can be obtained from a fit of ${f_\nu } \propto {\nu ^{ - ({\Gamma _{UV}} - 1)}}$ to u, g, r, i and z band, and the r-band apparent magnitude can also be used for calculating the UV-optical flux at (rest-frame) 2500 {\AA}, where $\left\langle {{\Gamma _{UV}}} \right\rangle  = 1.6$ are considered for the K-correction \citep{Richards2006}. In the same way, the observed 1.4 GHz flux is utilized to calculate radio flux at (rest-frame) 5 GHz by assuming ${a_r} =  - 0.5$. For X-ray fluxes of this sample, Galactic-absorption correction is performed by using PIMMs and obtain the unabsorbed flux density at observed-frame 2 keV, where a specifying Galactic column density and a power-law index in the X-Ray band $\left\langle {{\Gamma _X}} \right\rangle  = 1.6$ are considered, it can be used to determine band pass-corrected rest-frame 2 keV flux.

On the other hand, the blazars also can be used to check multi-band luminosity correlations, especially for FSRLQS. Recently Massaro present a multifrequency catalogue of blazars, named Roma-BZCAT which contains coordinates and multifrequency data of 3561 sources \citep{Massaro2009, Massaro2012, Massaro2015}. We match the Roma-BAZAT with the SDSS-DR16 quasar catalogue and obtain 407 FSRLQs with multi-wavelength coverage. At finally, this blazars sample and Type I quasars sample can be applied to investigate their luminosity correlations and measured the luminosity distances of these Type I RLQs.

In this paper, we only consider RLQs with $log R > 2$,  RIQs and RQQs satisfy $1< logR \le 2$ and $ logR \le 1$ respectively. Meanwhile, we employ parametric methods to test their multi-band luminosity correlation.

\section{THE RELATION BETWEEN X-RAY, UV-OPTICAL, AND RADIO LUMINOSITIES}\label{Sec:3}
\subsection{Insights from scatter plots} \label{Sec:3.1}
\begin{figure*}[htpb]
\begin{center}
\begin{minipage}[t]{0.49\textwidth}
\includegraphics[width=\linewidth,scale=1.00]{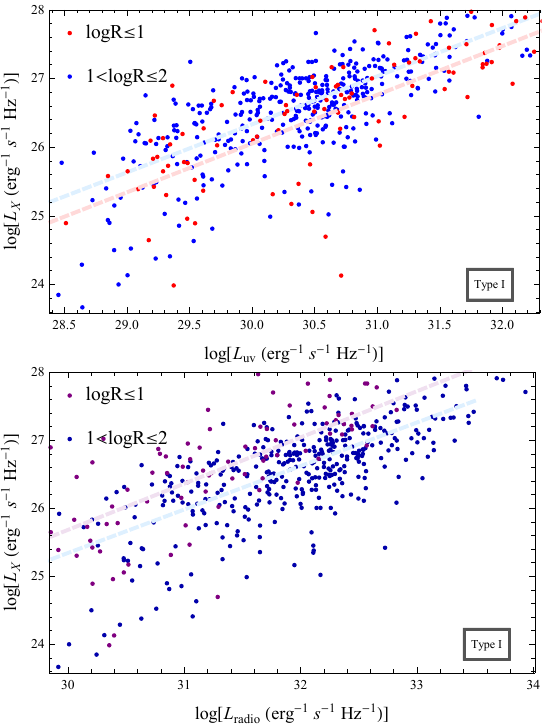}
\end{minipage}
\begin{minipage}[t]{0.49\textwidth}
\includegraphics[width=\linewidth,scale=1.00]{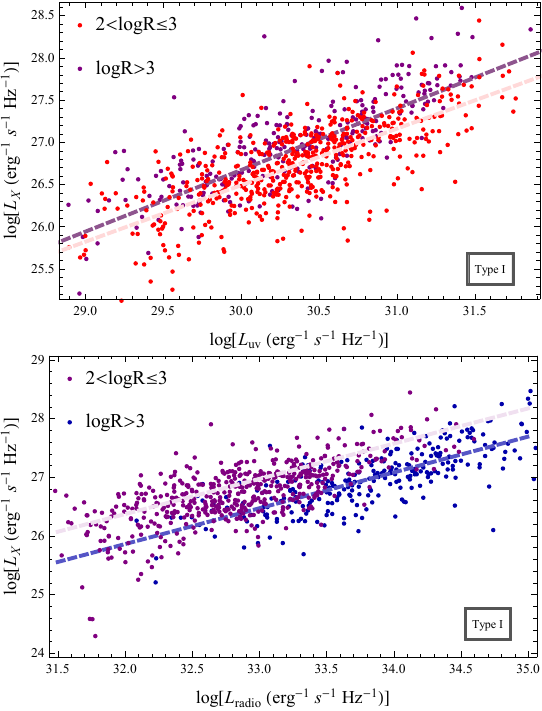}

\end{minipage}
\caption{Plot of $\log {\kern 1pt} {L_X} vs. \log {\kern 1pt} {L_{uv}} $ (upper panel) and $\log {\kern 1pt} {L_X} vs. \log {\kern 1pt} {L_{radio}} $ (lower panel) for Type I quasars with different $log R$, the dotted line represents the theoretical values of $\log {\kern 1pt} {L_X}$ luminosity from the linear relation together with the best fitting values of parameters.}
\label{fig:1}
\end{center}
\end{figure*}

\begin{figure}[htpb]
\centering
\includegraphics[width=\linewidth,scale=1.00]{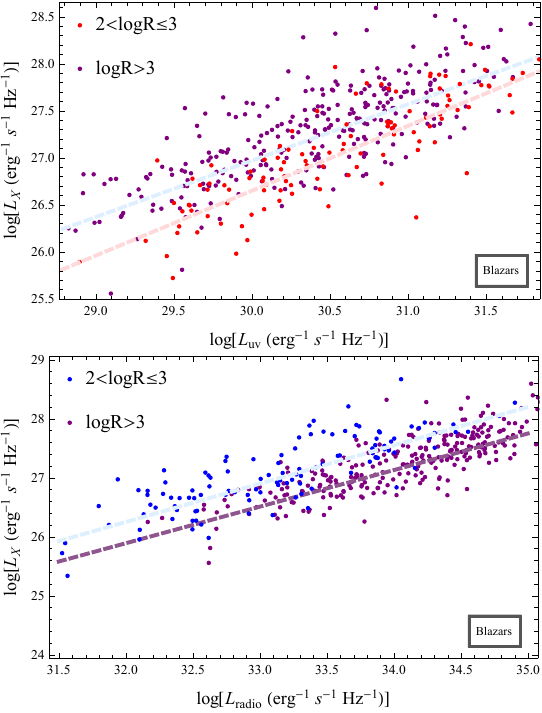}
\caption{Plot of $\log {\kern 1pt} {L_X} vs. \log {\kern 1pt} {L_{uv}} $ (upper panel) and $\log {\kern 1pt} {L_X} vs. \log {\kern 1pt} {L_{radio}} $ (lower panel) for blazars (FSRLQs) with different $log R$, the dotted line represents the theoretical values of $\log {\kern 1pt} {L_X}$ luminosity from the linear relation together with the best fitting values of parameters.}
\label{fig:2}
\end{figure}

\begin{figure}[htpb]
\centering
\includegraphics[width=\linewidth,scale=1.00]{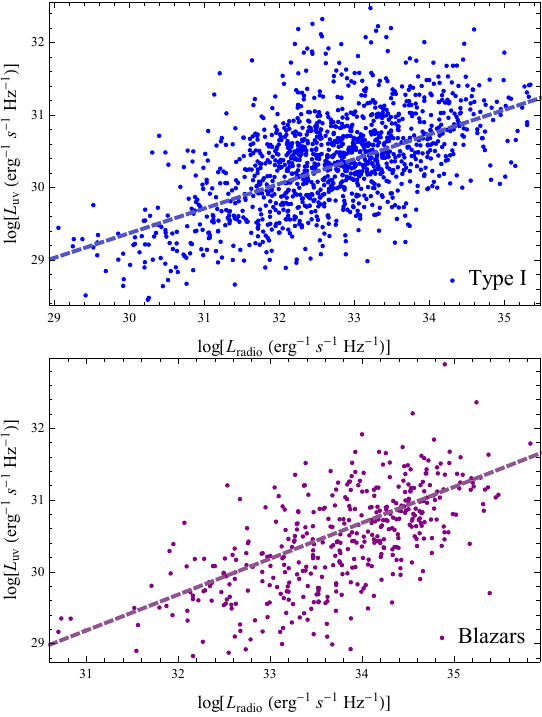}
\caption{Plot of $\log {\kern 1pt} {L_{uv}} vs. \log {\kern 1pt} {L_{radio}} $ for Type I quasars (upper panel) and blazars (lower panel), the dotted line represents the theoretical values of $\log {\kern 1pt} {L_X}$ luminosity from the linear relation together with the best fitting values of parameters.}
\label{fig:3}
\end{figure}

We firstly plot the ${L_X} - {L_{uv}}$ and ${L_X} - {L_{radio}}$ plane for Type I quasars and blazars (FSRLQs), as shown in Fig. \ref{fig:1} \ref{fig:2}, the luminosities ${L_\lambda }(2500{\AA})$ have been obtained from the measured fluxes assuming $\Lambda CDM$ cosmology $({\Omega _m} = 0.3,{\kern 1pt} {\kern 1pt} {H_0} = 70{\kern 1pt} km{\kern 1pt} {\kern 1pt} {s^{ - 1}}{\kern 1pt} Mp{c^{ - 1}})$. Meanwhile, we fit the linear relation to the data and obtain the theoretical values of X-Ray luminosity from the best fitting values of parameters. The upper left panel of Fig.\ref{fig:1} illustrates the ${L_X} - {L_{uv}}$ plane for Type I quasars with $logR \le 1$ (RQQs) and $1< logR \le 2$ (RIQs), and the dotted line represents the theoretical values of X-Ray luminosity from the linear relation together with the best fitting values of parameters, which implies that the X-ray luminosity of RQQs and RIQs is related to UV-optical luminosity and originates from the inverse compton scattering. The lower left panel of Fig.\ref{fig:1} shows the ${L_X} - {L_{radio}}$ plane for Type I RQQs and RIQs, and the dotted line represents the theoretical values, which also indicates that X-ray luminosity of RQQs is indirectly correlated with radio luminosity because of the connection between UV/optical and radio luminosity from Fig.\ref{fig:3} (${L_{uv}} - {L_{radio}}$ plane). The indirect relation between X-Ray and radio luminosity in RQQs will also be discussed in Sec \ref{Sec:3.4}.

Likewise, the ${L_X} - {L_{uv}}$ plane of Type I RLQs ($log R>2$) is shown in the upper right panel of Fig.\ref{fig:1}, which suggests that X-ray luminosity of RLQs is correlate with UV-optical luminosity. For the ${L_X} - {L_{radio}}$ plane of Type I RLQs illustrated in the lower right panel of Fig.\ref{fig:1}, whether or not the X-ray luminosity of RLQs is indirectly or directly related to radio luminosity will be discussed in Sec \ref{Sec:3.4}.

\begin{figure}[htpb]
\centering
\includegraphics[width=\linewidth,scale=1.00]{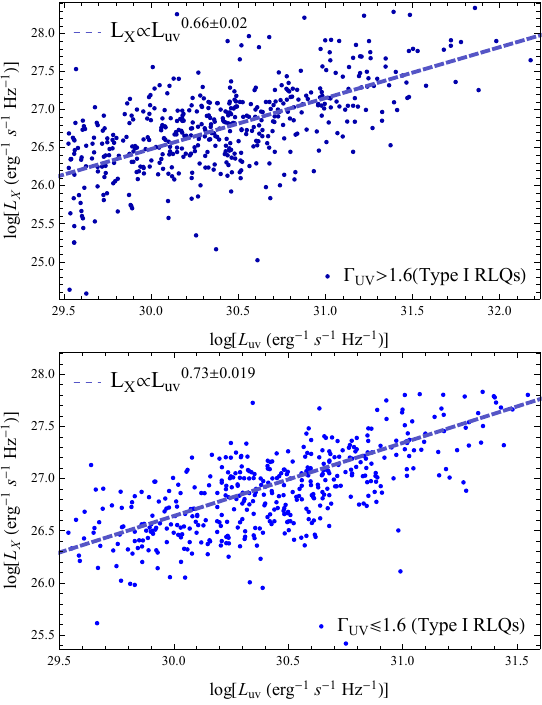}
\caption{Plot of $\log {\kern 1pt} {L_X} vs. \log {\kern 1pt} {L_{uv}} $ for Type I quasars with ${\Gamma _{UV}}> 1.6$ (upper panel) and ${\Gamma _{UV}}\le 1.6$ (lower panel), the dotted line represents the theoretical values of $\log {\kern 1pt} {L_X}$ luminosity from the linear relation together with the best fitting values of parameters.}
\label{fig:4}
\end{figure}

The ${L_X} - {L_{uv}}$ plane and ${L_X} - {L_{radio}}$ of blazars (FSRLQs) are illustrated in the upper and lower panel of Fig.\ref{fig:2}, which similarly implies that X-ray luminosity of RLQs is related to UV-optical luminosity. On the other hand, we compare the correlation between X-Ray luminosity and UV-optical luminosity of Type I quasars with UV-optical power-law ${\Gamma _{UV}}\le 1.6$ and ${\Gamma _{UV}}> 1.6$, which is shown in Fig.\ref{fig:4}. We will further discuss it in Sec \ref{Sec:3.4}.

\subsection{Models constraints from Type I quasars}

We apply various parameterization methods to test the multi-band luminosity correlation of quasars, which involve different physical mechanisms. The most common parametric equation is \citep{Tananbaum1983,Worrall1987,Miller2010,Zhu2020}
 \begin{eqnarray}\label{eq:1}
\begin{array}{l}
Model\\
I:\log {L_X} = \alpha  + {\gamma _{uv}}\log {L_{uv}} + \gamma _{radio}'\log {L_{radio}},
\end{array}
\end{eqnarray}

The above equation can become the relation ${L_X} \propto L_{uv}^{{\gamma _{uv}}}L_{radio}^{\gamma _{radio}'}$, which concerns that X-ray luminosity is related to both UV-optical luminosity and radio luminosity. Using formula $L = 4\pi {D_L}^2F$ in (\ref{eq:1}), we get
\begin{eqnarray}
\begin{array}{l}
\log {F_X} = \Phi ({F_{UV}},{F_{radio}},{D_L})\\
{\kern 1pt} {\kern 1pt} {\kern 1pt} {\kern 1pt} {\kern 1pt} {\kern 1pt} {\kern 1pt} {\kern 1pt} {\kern 1pt} {\kern 1pt} {\kern 1pt} {\kern 1pt} {\kern 1pt} {\kern 1pt} {\kern 1pt} {\kern 1pt} {\kern 1pt} {\kern 1pt} {\kern 1pt} {\kern 1pt} {\kern 1pt} {\kern 1pt} {\kern 1pt} {\kern 1pt} {\kern 1pt} {\kern 1pt} {\kern 1pt} {\kern 1pt} {\kern 1pt} {\kern 1pt} {\kern 1pt} {\kern 1pt} {\kern 1pt} {\kern 1pt}  = \alpha  + {\gamma _{uv}}\log {F_{UV}} + \gamma _{radio}'\log {F_{radio}}\\
{\kern 1pt} {\kern 1pt} {\kern 1pt} {\kern 1pt} {\kern 1pt} {\kern 1pt} {\kern 1pt} {\kern 1pt} {\kern 1pt} {\kern 1pt} {\kern 1pt} {\kern 1pt} {\kern 1pt} {\kern 1pt} {\kern 1pt} {\kern 1pt} {\kern 1pt} {\kern 1pt} {\kern 1pt} {\kern 1pt} {\kern 1pt} {\kern 1pt} {\kern 1pt} {\kern 1pt} {\kern 1pt} {\kern 1pt} {\kern 1pt} {\kern 1pt} {\kern 1pt} {\kern 1pt} {\kern 1pt} {\kern 1pt} {\kern 1pt} {\kern 1pt}  + ({\gamma _{uv}} + \gamma _{radio}' - 1)\log (4\pi {D_L}^2),
\end{array}\label{eq2}
\end{eqnarray}
where ${F_{X}}$, ${F_{UV}}$ and ${F_{radio}}$ are measured at (rest-frame) $2 keV$, $2500 Å$ and $5 GHz$, $D_L$ is the luminosity distance, which can be obtained by the integral formula of $D_L-z$ relation. This equation can be effectively used for testing  X-ray luminosity correlation for RLQs and RQQs.

The second model is considered that X-ray luminosity is only correlated with UV-optical luminosity, and its parametric form is \citep{Miller2010, Bisogni2021}
 \begin{eqnarray}\label{eq3}
IV:{\kern 1pt} {\kern 1pt} {\kern 1pt} {\kern 1pt} {\kern 1pt} {\kern 1pt} \log {L_X} = \alpha  + {\gamma _{uv}}\log {L_{uv}},
\end{eqnarray}

We can also consider other model as
 \begin{eqnarray}\label{eq4}
III:{\kern 1pt} {\kern 1pt} {\kern 1pt} {\kern 1pt} {\kern 1pt} {\kern 1pt} \log {L_X} = \alpha  + {\gamma _{radio}'}\log {L_{radio}},
\end{eqnarray}

Model II and Model III can become the relation ${L_X} \propto L_{uv}^{{\gamma _{uv}}}$ and ${L_X} \propto L_{radio}^{\gamma _{radio}'}$. The above two models refer that X-ray luminosity is only correlative with  UV-optical or radio luminosity.

In the same way, from equations (\ref{eq3}) and (\ref{eq4}), we can get X-ray flux ${F_{X}}$ as the function of ${F_{UV}}$, ${F_{radio}}$ and $D_L$, which can be used to test  X-ray luminosity relations.


We fit the three parametric models by minimizing a likelihood function consisting of a modified ${\chi ^2}$ function based on MCMC, allowing for an intrinsic dispersion $\sigma $ \citep{Risaliti2015}
\begin{large}
 \begin{eqnarray}\label{eq6}
\begin{array}{l}
 - 2\ln L = \sum\limits_{i = 1}^N {\left\{ {\frac{{{{[\log {{({F_X})}_i} - \Phi {{({F_{UV}},{F_{radio}},{D_L})}_i}]}^2}}}{{s_i^2}}} \right\}} \\
{\kern 1pt} {\kern 1pt} {\kern 1pt} {\kern 1pt} {\kern 1pt} {\kern 1pt} {\kern 1pt} {\kern 1pt} {\kern 1pt} {\kern 1pt} {\kern 1pt} {\kern 1pt} {\kern 1pt} {\kern 1pt} {\kern 1pt} {\kern 1pt} {\kern 1pt} {\kern 1pt} {\kern 1pt} {\kern 1pt} {\kern 1pt} {\kern 1pt} {\kern 1pt} {\kern 1pt} {\kern 1pt} {\kern 1pt} {\kern 1pt} {\kern 1pt} {\kern 1pt} {\kern 1pt} {\kern 1pt} {\kern 1pt} {\kern 1pt} {\kern 1pt} {\kern 1pt} {\kern 1pt} {\kern 1pt} {\kern 1pt}  + \sum\limits_{i = 1}^N {\ln (2\pi s_i^2)} ,
\end{array}
\end{eqnarray}
\end{large}
where $\Phi ({F_{UV}},{F_{radio}},{D_L})$ is given by equation (\ref{eq2}), and $s_i^2 = \sigma _i^2(\log {F_X}) + \gamma _{uv}^2\sigma _i^2(\log {F_{UV}}) + {\delta ^2} $, $\delta$ is the intrinsic dispersion, which can be fitted as a free parameter and $\delta$ is usually much larger than the measurement error.

The Hubble constant ${H_0}$ is degenerate with the parameters $\alpha $ when fitting equation (\ref{eq2}), we fix ${H_0} = 70{\kern 1pt} {\kern 1pt} km{\kern 1pt} {\kern 1pt} {\kern 1pt} {s^{ - 1}}{\kern 1pt} Mp{c^{ - 1}}$ \citep{Reid2019,Aghanim2020}. If we want to better test X-ray luminosity relations and further select the optimal model, we should not fix ${\Omega _m}$. Therefore, we fit the three models to Type I quasars without fixing ${\Omega _m}$ and seek the best model.

We adopt the maximum likelihood function (equation (\ref{eq6})) based on MCMC to constrain three models, the model fitting results for Type I quasars are illustrated in Table \ref{tab:1}. Meanwhile, we fit Model I to Type I quasars with ${\Gamma _{UV}}\le 1.6$ and ${\Gamma _{UV}}> 1.6$, as well as the objects with the X-Ray power-law index ${\Gamma _{X}}\le 1.6$ and ${\Gamma _{X}}> 1.6$, ${\Gamma _{X}}$ can be obtained from a fit of ${f_\nu } \propto {\nu ^{ - ({\Gamma _{X}} - 1)}}$ to their X-Ray fluxes(0.2-0.5, 0.5-1, 1-2, 4-4.5, 4.5-12 keV for XMM-Newtom and 0.5-1.2, 1.2-2, 2-7 keV for Chandra), the statistical results can be used to test whether there are different, which are also shown in Table \ref{tab:1}.

\begin{table*}[htpb]
\setlength{\tabcolsep}{2mm}
 \centering
  \caption{ Model fitting results for Type I RLQs and RQQs }
  \label{tab:1}
  \vspace{0.3cm}
  \begin{tabular}{@{}cccccccc@{}}
   \hline
  Model&Sample&$\alpha $&${\gamma _{uv}} $&${\gamma _{radio}'}$&$\delta $&${\Omega _m}$&$- 2In{L_{\max }}/N$\\
    \uppercase\expandafter{\romannumeral1}
   &$log R \le1$&4.12$\pm$1.47    &0.323$\pm$0.038&  0.4$\pm$0.075 & 0.65$\pm$0.049 &0.514$\pm$0.078 &185/97\\
   &$1 < log R \le 2$&4.12$\pm$0.93    &0.31$\pm$0.029&  0.41$\pm$0.031 & 0.478$\pm$0.018 &0.314$\pm$0.068  &527/385\\
   &$2 < log R \le 3$&5.09$\pm$0.652    &0.442$\pm$0.028&  0.253$\pm$0.025 & 0.281$\pm$0.009 &0.094$\pm$0.091  &144/492\\
   &$log R > 3$&4.5$\pm$0.8    &0.517$\pm$0.044&  0.202$\pm$0.028 & 0.3$\pm$0.014 &0.195$\pm$0.025  &89/218\\
   &$log R > 2;{\kern 1pt} {\Gamma _{UV}}  \le  1.6$&4.67$\pm$0.386    &0.505$\pm$0.01&  0.209$\pm$0.01 & 0.259$\pm$0.008 &0.033$\pm$0.017  &65/472\\
   &$log R > 2;{\kern 1pt} {\Gamma _{UV}} > 1.6$&6.54$\pm$0.819    &0.388$\pm$0.03&  0.258$\pm$0.016 & 0.329$\pm$0.015 &0.134$\pm$0.023  &144/238\\
    &$log R > 2;{\kern 1pt} {\Gamma _{X}} \le 1.6$&4.84$\pm$0.75    &0.382$\pm$0.026&  0.31$\pm$0.04 & 0.337$\pm$0.013 &0.308$\pm$0.032  &264.7/407\\
    &$log R > 2;{\kern 1pt} {\Gamma _{X}} > 1.6$&8$\pm$0.52    &0.4$\pm$0.021&  0.2$\pm$0.02 & 0.29$\pm$0.013 &0.347$\pm$0.1  &89.3/256\\
   \hline

    \uppercase\expandafter{\romannumeral2}
 &$log R \le 1$&6.53$\pm$0.774    &0.652$\pm$0.025&  $-$ & 0.64$\pm$0.047 &0.545$\pm$0.108  &186/97\\
   &$1 < log R \le 2$&5.17$\pm$0.8    &0.7$\pm$0.026&  $-$ & 0.482$\pm$0.017 &0.205$\pm$0.061 &533/385\\
   &$2 < log R \le 3$&5.99$\pm$0.658    &0.687$\pm$0.022&  $-$ & 0.288$\pm$0.009 &0.035$\pm$0.019 &174/492\\
   &$log R > 3$&5.19$\pm$0.97    &0.719$\pm$0.032&  $-$ & 0.307$\pm$0.015 &0.188$\pm$0.079  &100/218\\
   \hline
       \uppercase\expandafter{\romannumeral3}
 &$log R \le 1$&6.08$\pm$0.813&  $-$  &0.652$\pm$0.026&    0.661$\pm$0.056 &0.4$\pm$0.191  &193/97\\
   &$1 < log R \le 2$&5.86$\pm$1&  $-$   &0.648$\pm$0.031&   0.493$\pm$0.019 &0.175$\pm$0.042 &549/385\\
   &$2 < log R \le 3$&7.8$\pm$0.933&  $-$  &0.578$\pm$0.029&    0.299$\pm$0.017 &0.213$\pm$0.054 &203/492\\
   &$log R > 3$&8.52$\pm$0.427&  $-$  &0.547$\pm$0.013&    0.325$\pm$0.016 &0.343$\pm$0.052  &127/218\\
   \hline

\end{tabular}
\end{table*}

\begin{table*}[htpb]
\setlength{\tabcolsep}{2mm}
\centerline{}
\caption{Measured properties of the 710 Type I RLQs, $DM$ are the distance modulus from a fit of the X-ray luminosity relation ${L_X} \propto L_{uv}^{{\gamma _{uv}}}L_{radio}^{\gamma _{radio}'}$ with $\Lambda CDM$ model, ${\sigma _{DM}}$ are their error.  ${\Gamma _{UV}}{\kern 1pt} or{\kern 1pt} {\kern 1pt} {\Gamma _X} =  - 99$ is used for cases of non-detection.}
\label{tab:2}
\vspace{0.3cm}
\begin{tabular}{ccccccccccc}
\hline
$SDSS{\kern 1pt} {\kern 1pt} {\kern 1pt} name $&$z $&${m_i} $&${F_{UV}}$&${F_X}$&${F_{radio}}$&$\log {\kern 1pt} {\kern 1pt} R$&${\Gamma _{UV}}$&${\Gamma _{X}}$&$DM$&${\sigma _{DM}}$\\
\hline
094334.00+463332.1&	3.216&	20.401$\pm$0.026&	-24.149$\pm$0.010&	-27.739$\pm$0&	-20.859&	3.29&	2.004&	-99&	47.798&	2.444\\
090237.33+010135.9&	3.12&	20.601$\pm$0.041&	-24.184$\pm$0.016&	-27.803$\pm$0.054&	-21.58&	2.603&	1.856&	0.921&		46.841&	2.407\\
104909.81+373759.0&	3.003&	18.075$\pm$0.082&	-23.167$\pm$0.033&	-27.049$\pm$0.008&	-20.833&	2.333&	3.769&	1.382&		45.854&	2.370\\
160421.77+432354.6&	2.408&	19.389$\pm$0.015&	-23.658$\pm$0.006&	-27.023$\pm$0.007&	-20.634&	3.023&	1.437&	1.388&	44.311&	2.297\\
084218.39+362504.2&	2.244&	19.265$\pm$0.021&	-23.597$\pm$0.008&	-27.014$\pm$0.022&	-20.81&	2.786&	1.853&	1.435&		44.124&	2.304\\

\hline

\end{tabular}

\end{table*}

\begin{table*}[htpb]
\footnotesize
\setlength{\tabcolsep}{0.5mm}
 \centering
  \caption{Fit results on model parameters for a combination of SNla and Type I RLQs}
  \label{tab:3}
  \vspace{0.3cm}
  \begin{tabular}{@{}cccccccccccc@{}}
  \hline
&$Sample $&$\alpha $&${\gamma _{uv}} $&$\gamma _{radio}'$&$\delta $&${\Omega _m}$&${w_0}$&${w_\alpha }$&$\chi _{Total}^2$/${\chi'} _{Total}^2$/N\\
   $\Lambda CDM$
      &RLQs   & 5.29$\pm$0.212   & 0.44$\pm$0.01& 0.25$\pm$0.012&   0.283$\pm$0.008 & 0.176$\pm$0.013 &$-$&$-$ &232.8/708/710\\
   &SN+RLQs   & 6.155$\pm$0.388   & 0.434$\pm$0.019& 0.227$\pm$0.01&   0.285$\pm$0.007 & 0.271$\pm$0.007 &$-$&$-$ &1272.4/1746.6/1758\\
&SN+RLQs (${\Gamma _{UV}}  \le  1.6$) & 5.6$\pm$0.339   & 0.442$\pm$0.013& 0.224$\pm$0.011&   0.261$\pm$0.008 & 0.273$\pm$0.008 &$-$&$-$ &1107.6/1506.6/1520\\
&SN+RLQs (${\Gamma _{UV}}  >  1.6$) & 4.89$\pm$0.557   & 0.437$\pm$0.025& 0.262$\pm$0.02&   0.33$\pm$0.015 & 0.273$\pm$0.007 &$-$&$-$ &1182.5/1271.1/1286\\

   \hline
      ${w_0}{w_a}CDM$
   &SN+RLQs & 5.68$\pm$0.255   & 0.434$\pm$0.009& 0.24$\pm$0.006&   0.285$\pm$0.008 & 0.298$\pm$0.012 &-1.107$\pm$0.044&0.412$\pm$0.366 &1269.2/1743.4/1758\\
&SN+RLQs (${\Gamma _{UV}}  \le  1.6$) & 5.95$\pm$0.358   & 0.453$\pm$0.021& 0.216$\pm$0.024&   0.261$\pm$0.008 & 0.29$\pm$0.016 &-1.09$\pm$0.04&0.41$\pm$0.335&1104.7/1503.7/1520\\
&SN+RLQs (${\Gamma _{UV}}  >  1.6$) & 5.25$\pm$0.389   & 0.426$\pm$0.019& 0.261$\pm$0.014&   0.33$\pm$0.016 & 0.286$\pm$0.027 &-1.126$\pm$0.075&0.757$\pm$0.306&1178.8/1267.3/1286\\

   \hline

\end{tabular}
\end{table*}

\subsection{Models analysis and comparison} \label{Sec:3.4}

We use BIC to seek an optimal model. The BIC is
 \begin{equation}\label{eq11}
BIC =  - 2\ln {L_{\max }} + k{\kern 1pt} {\kern 1pt} \ln {\kern 1pt} {\kern 1pt} N,
\end{equation}
where ${L_{\max }}$ is the maximum likelihood, $k$ is the number of free parameters of the model, and $N$ is the number of data points.

For model II, by comparing the results in Tabel \ref{tab:1} from different logR of Type I quasars, we find that the correlation between X-ray and UV-optical luminosity increases with logR. Similarly, for the model I, the statistical results imply that the correlation between X-ray and radio luminosity becomes stronger as the ratio of monochromatic luminosities logR increases. Meanwhile, Model II has the smallest BIC by comparing the results in Table \ref{tab:1} from fitting for different models to RQQs, which indicates that the X-ray luminosity of RQQs is not directly correlated with their radio luminosity, but there is an indirect relation between X-ray and radio luminosity because of the connection between UV-optical and radio luminosity from Fig. \ref{fig:3}.

For RLQs, BIC for Model I is far smaller than Model II and III, which implies X-ray luminosity of RLQs is not only connected with optical/UV luminosity but also directly related to radio luminosity. A possible reason for the luminosity correlations RLQs is that a fraction of the nuclear X-ray emission is directly or indirectly powered by the radio jet, the specific physical mechanism needs to be further understood.

As for RIQs, By comparing BIC in Table \ref{tab:1} from fitting for different models to RIQs, we find that Model I has the smallest BIC, which might indicate that there is a weak correlation between the X-ray and radio luminosity of RIQs. Furthermore, for the fitting results BIC, there is a difference between Type I quasars with ${\Gamma _{UV}}\le 1.6$ and ${\Gamma _{UV}}> 1.6$ for Model I, the same goes for ${\Gamma _{X}}\le 1.6$ and ${\Gamma _{X}}> 1.6$. The goodness of fit for ${\Gamma _{UV}}\le 1.6$ and ${\Gamma _{X}}> 1.6$ seem to be better.

\subsection{Analysis of the relation ${L_X} \propto L_{uv{\kern 1pt} {\kern 1pt} }^{{\gamma _{uv}}}L_{radio{\kern 1pt} {\kern 1pt} }^{\gamma _{radio}'}$  } \label{Sec:3.5}

\begin{figure}[!t]
\includegraphics[width=\linewidth,scale=1.00]{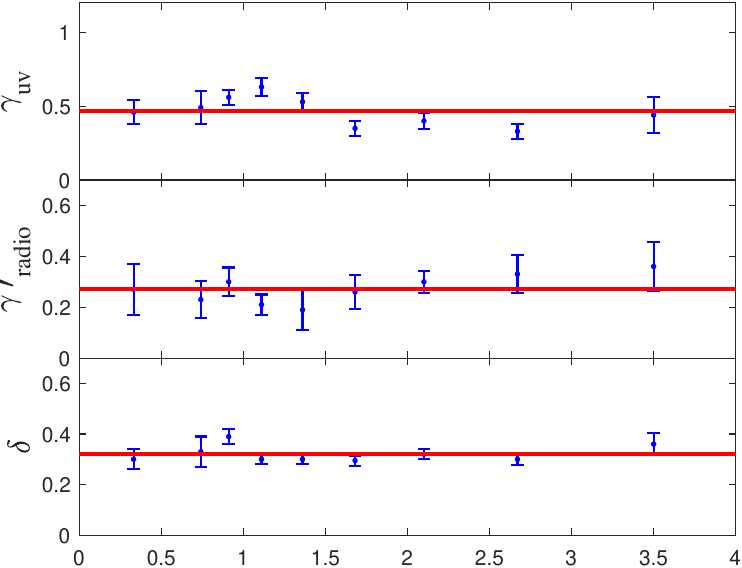}
\caption{${L_X} - {L_{uv}}/{L_{radio}}$ correlation in narrow redshift intervals. Blue points represent fitted values and errors of ${\gamma _{UV}}(z),{\kern 1pt} {\kern 1pt} \gamma _{radio}'(z),{\kern 1pt} {\kern 1pt} \sigma (z)$ at different redshift. The horizontal lines are their average values.}
\label{fig:5}
\end{figure}

We divide the Type I RLQs sample in several redshift bins, which can be used to verify if there is redshift dependence for luminosity relation. The redshift bin are $\Delta ({(1 + z)^{ - 1}}) = 0.05$. We apply the parametric model \citep{Risaliti2015}
\begin{large}
 \begin{eqnarray}\label{eq12}
\begin{array}{l}
\log {F_X} = \alpha (z) + {\gamma _{uv}}(z)\log {F_{UV}}\\
{\kern 1pt} {\kern 1pt} {\kern 1pt} {\kern 1pt} {\kern 1pt} {\kern 1pt} {\kern 1pt} {\kern 1pt} {\kern 1pt} {\kern 1pt} {\kern 1pt} {\kern 1pt} {\kern 1pt} {\kern 1pt} {\kern 1pt} {\kern 1pt} {\kern 1pt} {\kern 1pt} {\kern 1pt} {\kern 1pt} {\kern 1pt} {\kern 1pt} {\kern 1pt} {\kern 1pt} {\kern 1pt} {\kern 1pt} {\kern 1pt} {\kern 1pt} {\kern 1pt} {\kern 1pt} {\kern 1pt} {\kern 1pt}  + {\gamma _{radi{o^\prime }}}(z)\log {F_{radio}},
\end{array}
\end{eqnarray}
\end{large}
where $\alpha (z),{\kern 1pt} {\gamma _{uv}}(z),{\kern 1pt} \gamma _{radio}'(z)$ and the intrinsic dispersion $\delta (z)$ are free parameters. We fit equation (\ref{eq12}) to segmented Type I RLQs and check whether the X-Rays relation is dependent on the redshift. The fit results of ${\gamma _{uv}}(z),{\kern 1pt} {\kern 1pt} \gamma _{radio}'(z), and {\kern 1pt} {\kern 1pt} \delta (z)$ at different redshift are illustrated in Fig. \ref{fig:5}, which show that their values are no obvious evidence for any significant redshift evolution. The average values of parameters are $\left\langle {{\gamma _{uv}}} \right\rangle  = 0.47 \pm 0.1,{\kern 1pt} {\kern 1pt} {\kern 1pt} \left\langle {\gamma _{radio}'} \right\rangle  = 0.27 \pm 0.056.$

\section{A measure of luminosity distance  for Type I RLQs}\label{Sec:4}

Meanwhile, we measure the luminosity distance for Type I RLQs. From Model I, equation (\ref{eq2}) gives distance modulus as
 \begin{equation}\label{eq7}
DM = \frac{{5[\log {F_X} - {\gamma _{uv}}\log {F_{UV}} - \gamma _{radio}'{F_{radio}} - {\alpha '}]}}{{2({\gamma _{uv}} + \gamma _{radio}' - 1)}},
\end{equation}
where ${\alpha '} = \alpha  + ({\gamma _{uv}} + \gamma _{radio}' - 1)\log (4\pi )$.
 The formula of error is
 \begin{eqnarray}\label{eq8}
{\sigma _{DM}} = DM\sqrt {{{(\frac{{{\sigma _f}}}{f})}^2} + {{(\frac{{{\sigma _{{\gamma _{uv}}}}}}{\gamma })}^2} + {{(\frac{{{\sigma _{{\gamma _{radio}'}}}}}{\gamma })}^2}} .
\end{eqnarray}
where $f={\log {F_X} - {\gamma _{uv}}\log {F_{UV}} - {{\gamma '}_{radio}}{F_{radio}} - \alpha '}$, $\gamma  = {\gamma _{uv}} + {\gamma _{radio}'} - 1$, and ${\sigma _f}^2 = \sigma _i^2(\log {F_X}) + \gamma _{uv}^2\sigma _i^2(\log {F_{UV}}) + \sigma _{\alpha '}^2$. From equation (\ref{eq8}),  the uncertainty of the slope ${\gamma _{uv}}$ and $\gamma _{radio}'$  obviously influence the error of distance modulus for Type I RLQs.

\begin{figure}[htpb]
\centering
\includegraphics[width=\linewidth,scale=1.00]{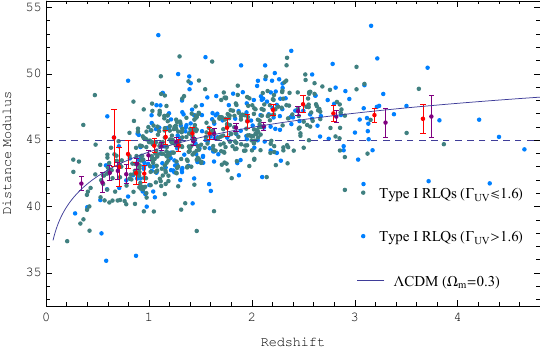}
\caption{The distance modulus of Type I quasars with ${\Gamma _{X}}\le 1.6$ and ${\Gamma _{X}}> 1.6$ from a fit of Equation (\ref{eq2}) when assuming $\Lambda CDM$ cosmology. The purple and red points are distance modulus averages in small redshift bins. The blue line shows a flat $\Lambda CDM$ model fit with ${\Omega _m} = 0.3$, the dotted line is the reference distance modulus and its value is 45.}
\label{fig:6}
\end{figure}

Fig \ref{fig:6} shows distance modulus of Type I quasars with ${\Gamma _{UV}}\le 1.6$ and ${\Gamma _{UV}}> 1.6$ from a fit of Model I when assuming $\Lambda CDM$ cosmology, and their averages in small redshift bins. Meanwhile the properties of 710 Type I quasars and their distance modulus are listed in Table \ref{tab:2}.

\section{The reconstruction of dark energy equation of state w(z)}\label{Sec:5}

Although the dark energy model can be used to effectively explain the accelerating expansion of the universe and the cosmic microwave background (CMB) anisotropies \citep{Riess1998, Amanullah2010, Betoule2014, Scolnic2018, Conley2010, Aghanim2020, Hu2002, Spergel2003, Ade2016, Aghanim2016}, the
origin and property of dark energy density and pressure are still unclear.

The research methods of dark energy include two kinds. One is to try to explain the physical origin of its density and pressure by constraining dark energy physical models  \citep{Peebles2003, Ratra1988, Li2004, Maziashvili2007, Amendola2000}. Understanding the physical nature of dark energy is important for our universe. Whether or not the dark energy is composed of Fermion pairs in a vacuum or Boson pairs, Higgs field. The order of magnitude for the strength of dark energy is far smaller than that the elementary particles needed when they were created in the very early Universe. The other method is to investigate whether or not the dark energy density evolves with time, this can be checked by reconstructing the dark energy equation of state $w(z)$ \citep{Linder2003, Maor2002}, which is independent of physical models. The high redshift observational data can better solve these problems.

 The reconstruction methods of the dark energy equation of state can be classed into parametric and non-parametric methods \citep{Huterer2003, Clarkson2010, Holsclaw2010, Seikel2012, Shafieloo2012, Crittenden2012, Zhao2012, Fu2019, Cao2017}. We apply Type I RLQs and SNla to reconstruct $w(z)$ by parametric method assuming X-ray luminosity relation Equation (\ref{eq2}), which can be used for testing the property of dark energy.

SNla Pantheon sample is a combination of data sources from the Sloan Digital Sky Survey (SDSS), the Pan-STARRS1 (PS1), SNLS, and various low-z and Hubble Space Telescope samples. There are 335 SNIa provided by SDSS \citep{Betoule2014,Gunn2006,Gunn1998,Sako2007,Sako2018}, and PS1 presented 279 SNla \citep{Scolnic2018}. The rest of the Pantheon sample are from the ${\rm{CfA1 - 4}}$, CSP, and Hubble Space Telescope (HST) SN surveys \citep{Amanullah2010, Conley2010}. This joint sample of 1048 SNIa is called the Pantheon sample.

The integral formula of ${D_L} - z$ relation in near flat space is given by
 \begin{equation}\label{eq13}
\begin{array}{l}
{D_L} = \frac{{1 + z}}{{{H_0}}}\int_0^z {d{z'}[{\Omega _m}{{(1 + {z'})}^3}} \\
{\kern 1pt} {\kern 1pt} {\kern 1pt} {\kern 1pt} {\kern 1pt} {\kern 1pt} {\kern 1pt} {\kern 1pt} {\kern 1pt} {\kern 1pt} {\kern 1pt} {\kern 1pt} {\kern 1pt} {\kern 1pt} {\kern 1pt} {\kern 1pt} {\kern 1pt} {\kern 1pt} {\kern 1pt} {\kern 1pt} {\kern 1pt} {\kern 1pt} {\kern 1pt} {\kern 1pt} {\kern 1pt} {\kern 1pt} {\kern 1pt} {\kern 1pt} {\kern 1pt} {\kern 1pt} {\kern 1pt} {\kern 1pt} {\kern 1pt} {\kern 1pt} {\kern 1pt} {\kern 1pt} {\kern 1pt} {\kern 1pt} {\kern 1pt} {\kern 1pt} {\kern 1pt} {\kern 1pt} {\kern 1pt} {\kern 1pt} {\kern 1pt} {\kern 1pt} {\kern 1pt} {\kern 1pt} {\kern 1pt} {\kern 1pt} {\kern 1pt}  + {\Omega _R}{(1 + {z'})^4} + \Omega _{DE}^{(0)}{{\mathop{\rm e}\nolimits} ^{\int_0^{{z'}} {\frac{{1 + w({z^{''}})}}{{1 + {z^{''}}}}d{z^{''}}} }}{]^{ - 1/2}}
\end{array}
\end{equation}
where ${{\Omega _R}}$ is radiation density. ${\Omega _{DE}^{(0)}}$ is the present dark energy density and satisfies $\Omega _{DE}^{(0)} = 1 - {\Omega _m}$ when ignoring ${{\Omega _R}}$, $w(z)$ is dark energy equation of state.
We choose ${w_0}{w_a}CDM$ model and the parametric form is
 \begin{equation}\label{eq13}
w(z) = {w_0} + {w_a}\frac{z}{{1 + z}}.
\end{equation}
Therefore dark energy density can be written as
 \begin{equation}\label{eq14}
{\Omega _{DE}}(z) = \Omega _{DE}^{(0)}{(1 + z)^{3(1 + {w_0} + {w_a})}}\exp [ - 3{w_a}z/(1 + z)].
\end{equation}

We fit ${{\rm{w}}_0}{{\rm{w}}_a}{\rm{CDM}}$ model parameters to Type I RLQs and SNla by minimizing $\chi _{Total}^2$, the $\chi _{Total}^2$ is
 \begin{eqnarray}\label{eq15}
\chi _{Total}^2 =  - 2\ln {L^{RLQs}} + \chi _{SN}^2,
\end{eqnarray}
where $ - 2\ln {L^{RLQs}}$ is given by equation (\ref{eq6}), and $\chi _{SN}^2$ can be expressed as
 \begin{equation}\label{eq16}
\chi _{SN}^2 = \Delta {\mu ^T}C_{{\mu _{ob}}}^{ - 1}\Delta \mu ,
\end{equation}
where $\Delta \mu  = \mu  - {\mu _{th}}$. ${C_\mu }$ is the covariance matrix of the distance modulus $\mu $.
Another function is ${\chi'} _{Total}^2$, which satisfies
\begin{equation}
{\chi'} _{Total}^2 = \chi _{RLQs}^2 + \chi _{SN}^2,
\end{equation}
and $\chi _{RLQs}^2 =  - 2\ln {L^{RLQs}} - \sum\limits_{i = 1}^N {\ln (2\pi s_i^2)} $.

\begin{figure}[htpb]
\includegraphics[width=\linewidth]{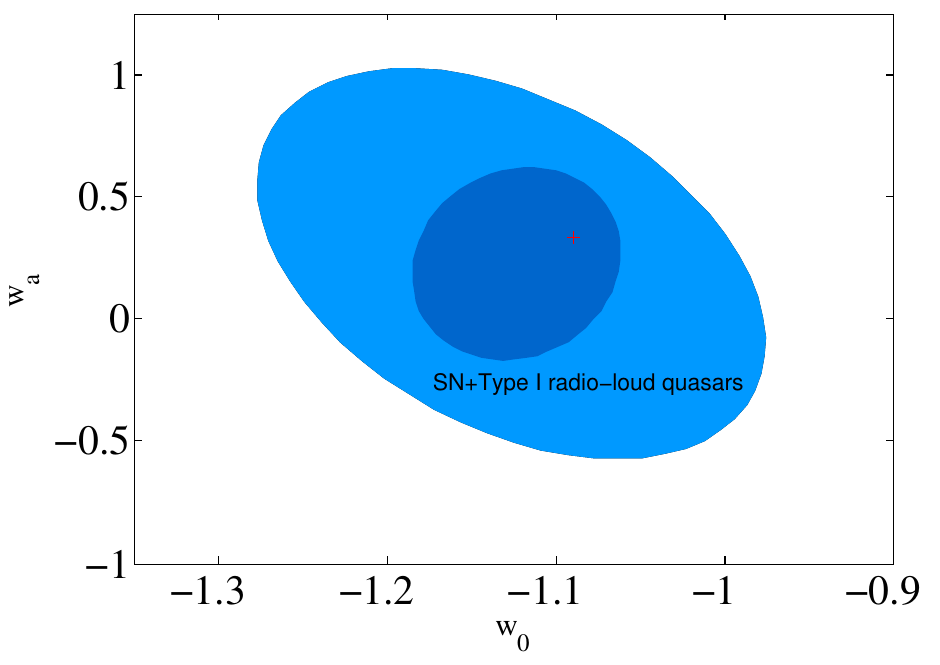}
\caption{$68\%$ and $95\%$ contours for ${w_0}$ and ${w_a}$ from a fit of the X-ray luminosity relation ${L_X} \propto L_{uv}^{{\gamma _{uv}}}L_{radio}^{\gamma _{radio}'}$ with $\Lambda CDM$ and ${w_0}{w_a}CDM$ model to a combination of SNla and Type I RLQs. The + dot in the responding color represents the best fitting values for ${w_0}$, ${w_a}$.}
\label{fig:7}
\end{figure}

We adopt equation (\ref{eq15}) to constrain model parameters, and fit results are illustrated in table \ref{tab:3}, ${w_0}{w_a}CDM$ has better goodness of fit than $\Lambda CDM$, and $\Delta \chi _{Total}^2$ is improved by $-3.2$, it implies $\Lambda CDM$ model is in tension with Type I RLQs at $\sim 1.5\sigma $, which is consistent with the results from the distance measurement using Baldwin effect of quasars \citep{Huang2022, Huang2023}. Meanwhile fig \ref{fig:7} shows  $68\%$ and $95\%$ contours for ${w_0}$ and ${w_a}$ from a fit of the X-ray luminosity relation ${L_X} \propto L_{uv}^{{\gamma _{uv}}}L_{radio}^{\gamma _{radio}'}$ and ${{\rm{w}}_0}{{\rm{w}}_a}{\rm{CDM}}$ model to a combination of SNla and Type I RLQs.

\section{Summary}\label{Sec:6}

The investigation of X-ray luminosity correlation for RQQs and RLQs could make us understand more of their physical mechanism. We obtain a new sample of 1192 Type I quasars with the UV-optical, radio and X-ray wavebands coverage by combining \citet{Huang2022} and other matching data of SDSS-DR16 with FIRST, XMM–Newton, and Chandra Source Catalog, and a sample of 407 flat-spectrum radio-loud quasars (FSRLQs) of blazars from the Roma-BZCAT. Firstly, we apply three parametric methods to test the correlation between X-ray, UV-optical, and radio luminosity. The statistical results indicate that the X-ray luminosity of RQQs is correlated with their UV-optical luminosity, which also can be considered that the X-ray luminosity of RQQs is indirectly correlated with radio luminosity because of the connection between UV-optical and radio luminosity.

Meanwhile, data suggest that the correlation between X-ray and UV-optical luminosity increases with the ratio of monochromatic luminosities $log R$, Similarly, the correlation between X-ray and radio luminosity also becomes stronger as $log R$ increases. For RLQs, the results imply that the X-ray luminosity of RLQs is not only connected with optical/UV luminosity but also directly related to radio luminosity. A possible reason for the luminosity correlations RLQs is that a fraction of the nuclear X-ray emission is directly or indirectly powered by the radio jet. In addition, we compare the results from Type I quasars with ${\Gamma _{UV}}\le 1.6$ and ${\Gamma _{UV}}> 1.6$,  as well as ${\Gamma _{X}}\le 1.6$ and ${\Gamma _{X}}> 1.6$ using a fit of X-ray luminosity relation ${L_X} \propto L_{uv}^{{\gamma _{uv}}}L_{radio}^{\gamma _{radio}'}$, the goodness of fit for ${\Gamma _{UV}}\le 1.6$ and ${\Gamma _{X}}> 1.6$ seem to be better.

Secondly, We divide the Type I RLQs sample into discrete redshift bins and combine a special model, which can be applied to describe if there is a redshift evolution of X-ray luminosity relation ${L_X} \propto L_{uv}^{{\gamma _{uv}}}L_{radio}^{\gamma _{radio}'}$, the fit results show the model parameters approach to the constant, which indicates there is not an obvious redshift dependence for ${L_X} \propto L_{uv}^{{\gamma _{uv}}}L_{radio}^{\gamma _{radio}'}$.

Finally, we obtain the luminosity distance of 710 Type I RLQs from a fit of X-ray luminosity relation ${L_X} \propto L_{uv}^{{\gamma _{uv}}}L_{radio}^{\gamma _{radio}'}$ when assuming $\Lambda CDM$ cosmology, and use a joint of SNla and Type I RLQs sample to reconstruct the dark energy equation of state $w(z)$ by parametric method and test the nature of dark energy. The data suggests ${w_0}{w_a}CDM$ model is superior to cosmological constant $\Lambda CDM$ model at $\sim 1.5\sigma $.

In the future, we will cross-correlate the Dark Energy Spectroscopic Instrument (DESI) quasar catalogs with the XMM-Newton, Chandra archives, and radio surveys. We expect to obtain more quasars with multi-wavelength coverage and high redshift ($z>3$) objects, which can be used to investigate their multi-band luminosity correlations. Meanwhile, the high redshift observational data can better test the properties of dark energy, it will determine the future of the universe, whether the universe keeps expanding or shifts from expansion to contraction. It will similarly determine the future of humanity.

\vspace{1cm}
\bibliographystyle{apsrev4-2}
\bibliography{bib}


\end{document}